\documentstyle[pra,aps,preprint]{revtex}
\tightenlines
\begin{document}
\title{Eigenvalue spectrum for single particle in a spheroidal cavity: A Semiclassical approach}
\author{Sham S. Malik$^{*,\dagger}$, A. K. Jain$^\dagger$ and S. R. Jain$^{**}$}
\address{$^*$Department of Physics, G.N.D. University, Amritsar-143005, India\\$^\dagger$Department of Physics, Indian Institute of Technology, Roorkee-247667, India\\$^{**}$Nuclear Physics Division, B.A.R.C., Mumbai-400085, India}
\maketitle
\begin{abstract}
Following the semiclassical formalism of Strutinsky et
al.\cite{str1977}, we have obtained the complete eigenvalue
spectrum for a particle enclosed in an infinitely high spheroidal
cavity. Our spheroidal trace formula also reproduces the results
of a spherical billiard in the limit $\eta\rightarrow1.0$.
Inclusion of repetition of each family of the orbits with
reference to the largest one significantly improves the
eigenvalues of sphere and an exact comparison with the quantum
mechanical results is observed upto the second decimal place for
$kR_{0}\geq{7}$. The contributions of the equatorial, the planar
(in the axis of symmetry plane) and the non-planar(3-Dimensional)
orbits are obtained from the same trace formula by using the
appropriate conditions. The resulting eigenvalues compare very
well with the quantum mechanical eigenvalues at normal
deformation. It is interesting that the partial sum of equatorial
orbits leads to eigenvalues with maximum angular momentum
projection, while the summing of planar orbits leads to
eigenvalues with $L_z=0$ except for L=1. The remaining quantum
mechanical eigenvalues are observed to arise from the
3-dimensional(3D) orbits. Very few spurious eigenvalues arise in
these partial sums. This result establishes the important role of
3D orbits even at normal deformations.
\end{abstract}


\section{Introduction}
Nuclei display a wide variety of phenomena, which originate from
single particle to collective motion and their interplay. A basic
framework for discussing these phenomena has been provided by the
shell model, which focuses upon the shell structure in the single
particle energy spectrum. Evolution of shell structure with
particle numbers, nuclear shapes and angular momenta has been a
central theme of nuclear structure physics for the past three
decades \cite{jain1998}. It has been the main guiding factor for a
number of theoretical predictions and explanations of several
nuclear phenomena such as fission isomers, shape coexistence,
superheavy elements and superdeformation. Strutinsky's shell
correction method \cite{str1967} has been a central approach to
the calculations of potential energy surfaces
\cite{bra1972,bjo1980}, which is a key to understand these
phenomena. The method, also known as the macroscopic plus
microscopic approach, obtains the fluctuating part of the level
density (the shell correction term) arising from the quantal shell
correction energy, which together with the liquid drop term gives
the total energy. The energy landscape so obtained have played an
important role in predicting phenomena like the occurrence of
superheavy elements \cite{hof2000} and superdeformation
\cite{bak1995}.

   Parallel to the Strutinsky' method, a new and radically different understanding of shell structure has evolved in recent times, which is based on the closed periodic paths in a given cavity \cite{gut1967,bal1970}. A semiclassical approach led Gutzwiller \cite{gut1990} to propose a trace formula for the fluctuating part of level density of a quantum system as
\begin{equation}
\delta{g}(E)={1\over{\pi\hbar}}\sum_{PO}{T_{PPO}\over{\sqrt{det{\mid{\tilde{M}_{PO}-I}\mid}}}}cos\bigl({S_{PO}\over{\hbar}}-\sigma_{PO}{\pi\over{2}}\bigr).
\end{equation}
The left hand side of equation (1) is a quantum mechanical quantity, whereas the right hand side contains only classical quantities. Here, $T_{PPO}$ represents the time period of the primitive periodic orbit or, the fundamental orbit and $\tilde{M}_{PO}$ is the monodromy stability matrix. The oscillations are controlled by action $S_{PO}$ of each periodic orbit and $\sigma_{PO}$ is the Maslov index. This trace formula is, however, applicable if all the periodic orbits are isolated and a summation is carried out over all periodic orbits.

    Strutinsky et al. \cite{str1975} generalized the Gutzwiller's trace formula to systems with continuous symmetries by explicitly taking into account the degeneracy of the classical motion. this method was successfully applied to phenomenological potentials used in nuclear physics \cite{str1977}, yielding a semiclassical interpretation of ground state deformation as a function of particle number.

    Later on, Frisk \cite{fri1990} presented calculations for the spheroidal cavities both in oblate and prolate region of deformations by using the method of Berry and Tabor \cite{ber1976}. For prolate deformation, his outcome is consistent with Strutinsky et al. and confirms the role of equatorial orbits in the shell structure at large deformation. In his calculations, the contribution of equatorial orbits were included from the formalism of Balian and Bloch \cite{bal1970}, as the Berry and Tabor formula was not applicable for such type of orbits. In a more recent work carried out by Magner et al \cite{mag1997}, the Jacobian of transformation was calculated by using the caustic method. Further in the spherical limit, it shows divergences. Therefore, the calculations in the spherical limit were done by the so called the bridge trace formula.

    The aim of the present paper is to bring the work of Strutinsky et al to a logical conclusion and to obtain the complete trace formula for spheroidal cavity, free from the above limitations. Here, we obtain the explicit eigenvalue spectrum for a particle enclosed in an infinitely deep spheroidal well and establish the importance of various types of orbits viz. the planar as well as the non-planar, 3D orbits. A comparison of the spectrum with the quantum mechanical eigenvalue spectrum \cite{mos1955} is made. Our main emphasis is on the role of 3D orbits in reproducing the eigenvalue spectrum even at normal deformation. Since these orbits are longer in length as compared to the planar orbits, it has been believed that 3D orbits have very little role in the shell structure and other properties. However our work conclusively demonstrates that the full eigenvalue spectrum cannot be obtained even at small deformation if 3D orbits are not included.

    It may be noted that the trace formula has also been successfully used in explaining the magic numbers of metal clusters \cite{ped1991}. The super shell structure predicted by Balian and Bloch \cite{bal1970} or, the beat pattern in the level density of a spherical billiard \cite{mal1999} can be understood very simply in terms of the interference of the shortest two orbits, namely, the square and the triangular orbits. This semiclassical approach has been extended by Brack et al \cite{bra11997} to provide a beautiful and more transparent explanation of the mass asymmetry in nuclear fission; the only physical input is the Fermi energy where the shell correction is minimum for the fission isomeric shape. Inclusion of the shortest orbits with constant action in a deformed cavity and the related pitchfork bifurcation was enough to reproduce this effect. More recently, this semiclassical approach has been used in \cite{mag2001} to obtain the bifurcation of 3D orbits, establishing its importance in the region of superdeformation by obtaining the gross shell structure.

    The trace formula for spheroidal cavity is given in Section II. In this formula, the bifurcation of 3D orbits begins to appear at specific minimum deformation and continues afterwards. Section III is devoted to results and discussion. The conclusions are summarized in Section IV. The derivation of Jacobian of transformation, which involves classical equations of motion, is given in the Appendix.

\section{Trace formula for spheroidal cavity}
    For the bound states of a single particle Hamiltonian $H$, the level density is generally defined as a sum of Dirac delta distributions,
\begin{equation}
g(E)=\sum_{i}\delta(E-E_{i}).
\end{equation}
This sum runs over all the eigen-energies $E_{i}$ (including degeneracies). It can be split into a smooth and an oscillating part:
\begin{equation}
g(E)=\tilde{g}(E)+\delta{g}(E).
\end{equation}
Following Strutinsky et al. \cite{str1977}, the oscillating component of single particle level density, $\delta{g}(E)$, in cylindrical coordinates $(\rho,\phi,z)$ is given by
\begin{equation}
\delta{g}(E)={1\over{\pi\hbar^2}}\sum_{\beta,m}f_{\beta,m}sin({S_{\beta,m}\over{\hbar}}+\alpha_{\beta,m}){\int\int}d{\rho}dz\sqrt{\mid{p_{\rho}\rho{J(p_{\rho}p_{z}t_{\beta,m};{\rho^\prime}{z^\prime}E)}}\mid}_{{\rho^\prime},{z^\prime}\rightarrow{\rho},z}.
\end{equation}
    The factor $f_{\beta,m}$ equals to 1 for the diametric orbits and 2 for other orbits like triangles, squares etc. The time period for the path from the initial point $\vec{r}$ to the final point $\vec{r^\prime}$ for energy E is defined as
\begin{equation}
t_{\beta}={\partial{S_{\beta}}(\vec{r},\vec{r^\prime},E)\over{\partial{E}}}.
\end{equation}
The quantity J in equation (4) is the Jacobian of transformation between two sets of classical quantities $(p_{\rho},p_{z},t_{\beta,m})$ and $({\rho^\prime},z^\prime,E)$, which are related by the classical equations of motion. Here, $\beta$ denotes the type of orbits and m is the number of repetitions of a given type of orbit.

    In this study, we have approximated the nuclear mean field by a spheroidal cavity with an infinitely high boundary, so that a particle makes a perfect reflection on striking the walls. The potential inside the well is assumed to be zero. The Hamiltonian for a single particle of mass M is expressed in terms of spheroidal coordinates $(u,v,\phi)$ and the conjugate momenta $(p_{u},p_{v},p_{\phi})$ as
\begin{equation}
H={{p_{u}^2+p_{v}^2}\over{2M{\zeta^2}(cosh^2{v}-sin^2{u})}}+{p_{\phi}^2\over{2M{\zeta^2}cos^2{u}sinh^2{v}}}
\end{equation}
Here, the parameter $\zeta$ (=$\sqrt{c^2-a^2}$; c and a being the semi-major and semi-minor axes, respectively)is fixed by the volume conservation condition and is given by
\begin{equation}
\zeta=R_{0}{\eta^{-{1\over{3}}}\sqrt{{\eta^2}-1}},
\end{equation}
where, $\eta (={c\over{a}})$ is the deformation parameter and $R_{0}$ is the radius of an equivalent sphere.

Equation (6) leads to the Hamilton Jacobi equation, which can be solved by using the separation of variable technique. Therefore, the classical dynamics of a particle is determined by the following three partial actions:
\begin{eqnarray}{\nonumber}
s_{u}=\oint{du{\sqrt{2M{\zeta^2}\bigl[\epsilon-E{sin^2{u}}-{l_{z}^2\over{2M{\zeta^2}cos^2{u}}}\bigr]}}}
\end{eqnarray}
\nonumber
\begin{eqnarray}{\nonumber}
s_{v}=\oint{dv{\sqrt{2M{\zeta^2}\bigl[E{cosh^2{v}}-\epsilon-{l_{z}^2\over{2M{\zeta^2}sinh^2{v}}}\bigr]}}}
\end{eqnarray}
\begin{equation}
s_{\phi}=2{\pi}l_{z}
\end{equation}
The constants of motion $\epsilon$ and $l_{z}$ are related to the turning points $-u_{max}, u_{max}, v_{min}, v_{0}$ as given in equations (19)-(21) of \cite{str1977}, which, in turn, are fixed by the following periodicity conditions:
\begin{equation}
{\omega_{u}\over{\omega_{v}}}={1\over{2}}\bigl[1-{F(\theta,\kappa)\over{F({\pi\over{2}},\kappa)}}\bigr]={n_{u}\over{n_{v}}}
\end{equation}
\begin{equation}
{\omega_{\phi}\over{\omega_{u}}}={n_{\phi}\over{n_{u}}}
\end{equation}
Here, $n_u, n_{v}$ and $n_{\phi}$ are co-prime integers, which define a family of periodic orbits. An explicit expression for ${\omega_{\phi}\over{\omega_{u}}}$ is given in \cite{str1977}. In equations (9) and (10), we have used the standard definitions of the elliptic integrals of the first and third kind with $\kappa$ and $\theta$ defined as
\begin{eqnarray}{\nonumber}
\kappa={sin{u_{max}}\over{cosh{v_{min}}}}
\end{eqnarray}
and
\begin{equation}
\theta=sin^{-1}\bigl({cosh{v_{min}}\over{cosh{v_{0}}}}\bigr)
\end{equation}
The total action in equation (4) is thus obtained as
\begin{equation}
S_{\beta}=n_{u}s_{u}+n_{v}s_{v}+n_{\phi}s_{\phi}=\sqrt{2ME}L_{\beta}
\end{equation}
where, $L_{\beta}$ is the length of the orbit. By using the transformation of coordinates, the radial $(\rho)$ and axial (z) coordinates are given by
\begin{eqnarray}{\nonumber}
\rho=\zeta{cos{u}sinh{v}}
\end{eqnarray}
\begin{equation}
z=\zeta{sin{u}cosh{v}}
\end{equation}
The corresponding canonically conjugate momenta $p_{\rho}$ and $p_z$ are
\begin{equation}
p_{\rho}=\sqrt{2ME}{[-Asin{u}sinh{v}+Bcos{u}cosh{v}]\over{cosh^2{v}-sin^2{u}}}
\end{equation}
\begin{equation}
p_z=\sqrt{2ME}{[Acos{u}cosh{v}+Bsin{u}sinh{v}]\over{cosh^2{v}-sin^2{u}}}
\end{equation}
where,
\begin{equation}
A=\sqrt{\sigma_{1}-sin^2{u}-{\sigma_{2}\over\cos^2{u}}}
\end{equation}
and
\begin{equation}
B=\sqrt{cosh^2{v}-\sigma_{1}-{\sigma_{2}\over\sinh^2{v}}}
\end{equation}
with $\sigma_{1}={\epsilon\over{E}}$ and $\sigma_{2}={l_{z}^2\over{2ME{\zeta^2}}}$. An explicit analytical expression for Jacobian of transformation, J, is derived in the Appendix. Here, we would like to mention that Strutinsky et al. \cite{str1977} have given only a rough estimate for J, while the same is obtained by Magner et al. \cite{mag1997}using the caustic method. Our approach for the calculations of J is simpler.

    Next, the maximum limit for the repetition parameter $m$ in equation (4) is obtained by considering that the longest periodic orbit, out of the permissible families $\beta$, traverses only once in the cavity. This fixes the repetition numbers of the individual families with reference to that of the largest one i.e.
\begin{equation}
 m_{max}={L_{largest}\over{L_{\beta}(n_{u},n_{v},n_{\phi})}}.
\end{equation}
 This technique is enough to remove the arbitrariness in the choice of $m_{max}$.

    In the spheroidal cavity, we have the planar as well as the 3D orbits. The planar orbits are further classified into two categories. One is in the plane perpendicular to the axis of symmetry (z=0) and the other is in the plane containing the symmetry axis $(l_{z}=0)$. The former correspond to the equatorial planar orbits, hereafter referred to as Type I. The later category is designated as Type II.

    Type I orbits have a lower degeneracy (K=1), while that of Type II is higher (K=2). The diametric orbit in the Type II family is isolated and has K=0. The 3D orbits have a degeneracy (K=2) same as that of Type II. We have included the effect of lower degeneracies in our calculations by dividing equation (4) by a factor $\sqrt{kR_{0}}$ and $kR_{0}$, respectively, for K=1 and K=0. Here, the wave number $k={\sqrt{2ME}\over\hbar}$.

    The orbits of Type I are determined by the condition z=0 and the equality ${\omega_{v}\over{\omega_{\phi}}}={n_{v}\over{n_{\phi}}}$. This condition can be obtained by using both equations (9) and (10). On the other hand Type II orbits are obtained from the periodicity condition (9) along with ${l_{z}}=0$.

    More important are the 3D orbits, which are characterized by the solutions $\kappa(n_{u},n_{v},n_{\phi})$ and $\theta(n_{u},n_{v},n_{\phi})$. These solutions exist for deformation $\eta\geq{\eta_{min}}$, where
\begin{equation}
\eta_{min}={sin({\pi{n_{\phi}}\over{n_{v}}})\over{sin({\pi{n_{u}}\over{n_{v}}})}}
\end{equation}
at which point $\kappa=0, \theta={\pi\over{2}}(1-2{n_{u}\over{n_{v}}})$ along with the conditions
\begin{equation}
v_{0}>{v_{min}}>0; ~~0<u_{max}<{\pi\over{2}}.
\end{equation}

    The Maslov index $\alpha_{\beta,m}$ in (4) plays a very crucial role in the periodic orbit sum as it decides the relative phase of the various terms in the summation. Following Creagh and Littlejohn \cite{cre1990} and Brack and Bhaduri \cite{bra1997}, we obtain the following Maslov indices for Type I, Type II and 3D orbits.\\
Type I: For diametric orbits
\begin{equation}
\alpha_{\beta,m}=-{3\over{2}}mn_{v}\pi+{\pi\over{4}};
\end{equation}
for other orbits
\begin{equation}
\alpha_{\beta,m}=-{3\over{2}}mn_{v}\pi-(mn_{\phi}-1)\pi+{\pi\over{4}}.
\end{equation}
Type II: For diametric orbits
\begin{equation}
\alpha_{\beta,m}=-{3\over{2}}mn_{v}\pi+{\pi\over{4}};
\end{equation}
for other orbits
\begin{equation}
\alpha_{\beta,m}=-{3\over{2}}mn_{v}\pi-(mn_{u}-1)\pi+{\pi\over{4}}.
\end{equation}
For 3D:
\begin{equation}
\alpha_{\beta,m}=-{3\over{2}}mn_{v}\pi-(mn_{u}-1)\pi-mn_{\phi}\pi+{\pi\over{4}}
\end{equation}

    Thus, the fluctuating part of the level density is fully established by including all the above mentioned terms in equation (4). The total level density $\delta{g}(E)$, is
\begin{equation}
\delta{g}(E)=\delta{g}^{Type I}+\delta{g}^{Type
II}+\delta{g}^{3D}.
\end{equation}

    Since the maximum contribution to the gross shell structure comes from the shortest periodic orbits, it has now become customary to carry out a smooth truncation of the contributions of the longer periodic orbits by folding the level density with a Gaussian function \cite{bra1997} of width ${\gamma\over{R_{0}}}$,
\begin{equation}
\delta{g}_{av}=\sum_{\beta,m}\delta{g}(E)exp[-({{\gamma}L_{\beta,m}\over{2R_{0}}})^2].
\end{equation}
The averaging width $\gamma$ is chosen to be larger than the mean spacing between the energy levels within a shell, but much smaller than the distance between the gross shells. This averaging ensures that all longer paths are strongly damped and only the shortest periodic orbits contribute to the oscillating part of the level density.

    Finally, the smooth part is obtained by using the Weyl's expansion \cite{bra1997} as given below:
\begin{equation}
\tilde{g}(E)={1\over{4{\pi^2}}}({2M\over{\hbar^2}})^{3\over{2}}\sqrt{E}V-{1\over{16\pi}}({2M\over{\hbar^2}})S+{1\over{12{\pi^2}}}({2M\over{\hbar^2}})^{1\over{2}}{C\over{\sqrt{E}}}.
\end{equation}
Here, V, S and C refer to the volume, surface area and radius of curvature,  of the cavity respectively. Following the geometrical procedure, we have obtained these terms for the spheroidal cavity and are given below:
\begin{equation}
V={4\over{3}}\pi{a^2}c,
\end{equation}
\begin{equation}
S=2\pi{a^2}\bigl[1+{sin^{-1}{f}\over{f{\sqrt{1-f^2}}}}\bigr],
\end{equation}
\begin{equation}
C={c\over{2}}[1+{{1-{f^2}}\over{2f}}ln{{1+f}\over{1-f}}],
\end{equation}
with $f={\sqrt{c^2-a^2}\over{c}}$.

\section{Results and Discussion}
    Firstly, we have obtained the semiclassical eigenvalue spectrum at $\eta=1.0001$ (in the limit of spherical shape). In our numerical calculations, we have varied $n_{\phi}$=1 to 200 and $n_{v}=2n_{\phi}$ to 200 for Type I and $n_{u}$=1 to 200 and $n_{v}=2n_{u}$ to 200 for Type II orbits. The parameter m is varied from 1 to $m_{max}$. Here, the convergence factor $\gamma$ is chosen as $0.02R_{0}$. The results of $g(E)$ vs. $kR_{0}$ arising from partial sums for Type I and II orbits are shown in Figs. 1 and 2, respectively. It is interesting enough to note that both these figures 1 and 2 show the same spectrum and are in good agreement with the quantum mechanical eigenvalues of spherical cavity. An exact comparison is seen upto second decimal place at $kR_{0}\geq7$. The combined spectrum of Type I and II orbits is shown in Fig. 3. These results confirm that our calculations for both Type I and Type II orbits give correct eigenvalues in the spherical limit. The partial sums over Type I and II orbits presented above lead us to similar results in the spherical limit as the two types of orbits become identical in this limit. It is well known that the trace formula calls for a simultaneous sum of all orbits. We, however, present the results of partial sums also in view of their direct relationship with eigenvalues belonging to distinct symmetry classes; this becomes more clear when a deformation is introduced.

    Here, we would like to mention that our present results are greatly improved in comparison to our earlier ones \cite{mal1999} due to the following reasons.
(i) The inclusion of the sum over repetition of the periodic
orbits, which has not been taken into account in our earlier
calculations. (ii) The last term in the Maslov indices, as given
in equations (21)-(24), has been replaced from ${3\pi\over{4}}$ to
${\pi\over{4}}$. The present spheroidal trace formula, which
diverges for $\eta$=1.0, still gives very good results in the
limit $\eta\rightarrow1.0$. We have separately obtained a trace
formula for the case of the spherical billiard by using the same
approach \cite{mal1999}. The trace formula of Magner et al
\cite{mag1997} also diverges in the spherical limit; as a result,
they have obtained their results in this limit by using the so
called bridge trace formula.

    As we increase the deformation from the spherical limit, interesting results emerge from our trace formula. Here, we discuss the results at $\eta=1.3$, a normal value of deformation in nuclei. Also, we again carry out partial sums as well as total sum. The convergence factor $ \gamma $ is taken to be 0. This choice ensures that no eigenvalue in the spectrum is missing. The variation of $(n_{u},n_{v},n_{\phi})$ for Type I and II orbits is kept same as in the spherical limit. For 3D orbits, we have varied $n_{u}$=1 to 200, $n_{v}=2n_{u}$ to 200 and $n_{\phi}$=1 to 100. The variation of $g(E)$ vs. $kR_{0}$ for each of the three families viz. Type I, II and 3D are shown in Figs. 4-6, respectively. On comparing these results with that of quantum mechanical one obtained from Ref. \cite{mos1955}, the following salient features emerge.\\
(i) All the quantum mechanical eigenvalues corresponding to maximum projection of the angular momentum i.e $L=L_{z}$ are seen in Fig. 4 that refers to Type I.
(ii)    The quantum mechanical values corresponding to $L>2N$ (N being the usual quantum number) and $L_{z}=0$ are visible in Fig. 5 that refers to Type II.\\
(iii) Rest of the quantum mechanical eigenvalues i.e.
$L\neq{L_{z}}$ are seen in our 3D plots, shown in Fig.6.

    In order to understand the origin of these remarkable results, let us once again consider the results shown for the above spherical case. The identical nature of the spectrum of Type I and II orbits, as shown in Figs. 1 and 2, respectively, reveals that periodic orbit theory can not separate the eigenvalues of Type I (refer to $L=L_{z}$) from that of Type II (refer to $L\neq{L_{z}}$). It may be remarked that 3D orbits do not exist in the spherical limit due to condition(19).  Quantum mechanics also shows the same characteristics. This shows that the outcome of periodic orbit theory is consistent with that of the quantum mechanics. The emergence of eigenvalues belonging to distinct symmetry sectors from the partial sums of distinct types of orbits is at once striking and beautiful.

    The increase in deformation from spherical limit to $\eta=1.3$, shows the complete separation of the eigenvalues of Type I from that of Type II and 3D orbits. Our quantum mechanical results (see Table 1) also show that  $L=L_{z}$ eigenvalues have no overlap with any of the eigenvalues of $L\neq{L_{z}}$. Here, no overlap means, the difference between these two sets is quite significant. This reveals that there is one to one correspondence between the two results.

    An interesting point is that all these spectral lines in Fig. 5, correspond to the lowest angular momentum projection quantum mechanical eigenvalues only. At this point our results differ from that of \cite{mag1997}. In their calculations of gross shell structure, they mentioned that Type II orbits have a higher degeneracy (K=2) and contribute more strongly at smaller and normal deformation. At smaller deformation, it is true and consistent with our results shown in Fig.2 (see the amplitudes in Fig.1 and 2). But at normal deformation, Type II orbits show only the lowest projection eigenvalues. Its role in comparison to the 3D orbits appear to be  much smaller. Since, 3D orbits are larger in length, so Magner et al have not seen its contribution in the gross shell structure.

    An explicit comparison of each of the three types of orbits is given in Table I. It is clear from comparison that the difference between eigenvalues of Type I orbits and that of quantum calculations is nearly constant ($\sim 0.26$). This difference may be due to choice of Maslov index and the less number of orbits in the sum. Here, we would like to mention that in all our present calculations, we have not changed the choice of the Maslov index.

    Finally, the complete spectrum at $\eta=1.3$ is shown in Fig. 7. As the Type I orbits have smaller degeneracy, so their contribution is not seen in the final plot. On the other hand, 3D orbits have higher degeneracy and occupy the larger volume at normal deformation.

\section{Conclusions}
A complete  trace formula for the level density of a spheroidal
cavity is presented which follows the Strutinsky's trace formula.
This formula reproduces the eigenvalues of sphere as we approach
the spherical limit and comparison with that of quantum mechanical
ones is found to be excellent upto second decimal place at
$kR_{0}\geq{7}$. The same formula works well for equatorial, axis
of symmetry plane and 3D orbits with appropriate boundary
conditions. Contrary to the general prescription, at  normal
deformation, the 3D orbits are found to be more important as
compared to the planar orbits in the axis of symmetry plane, while
equatorial orbits reproduce only the quantum mechanical
eigenvalues corresponding to maximum projection of angular
momentum. The overall eigenvalue spectrum compares reasonably well
with that of the quantum mechanical one. Our studies firmly
establish the expected, important role of the 3D orbits in the
appearance of the spectrum of a particle in a spheroidal well in
spite of the fact that they are longer in length.

Our studies suggest a way for a "semiclassical" shell model. We
have been able to identify each quantum level with a specific type
of periodic orbits. If we consider shell model built on these
orbits, we can fill up two fermions of opposite spin in each
state. Seen in this way, the magic numbers would appear at large
deformation only when we consider the filling of the states lying
in bunches. The bunching of levels would not be reproduced
correctly unless the 3D orbits are correctly placed.  Thus, the
shell structure may now be entirely understood in terms of
occupation of particles on periodic orbits instead of being on
energy levels. Also, the fact that each level is marked by a
periodic orbit of a certain kind, shows that the entire spectrum
of the spheroidal system has eigenfunctions which are scarred by
periodic orbits \cite{heller84}.

Such a trace formula can be envisaged to have a number of
applications from nuclei to atoms and metallic clusters.
Calculation of 3D plots of the deformation energy based on the
periodic orbit trace formula could successfully explain the onset
of mass asymmetry of fission in terms of few periodic orbits
\cite{bra11997}. A large number of studies have recently applied
this approach to the metal clusters \cite{mbc97,bbcmmr97}, and
quantum dots \cite{rplb96} with a good measure of success. It is,
therefore, expected that the formulation presented here will play
a useful role and lead to further applications in nuclei and
metallic clusters.

\section{Acknowledgment}
We acknowledge financial support from the Department of Atomic Energy (Govt. of India) and the Department of Science and Technology (Govt. of India).

\section{APPENDIX}
\title{Calculation of the Jacobian}
The Jacobian, J, of transformation is defined as
\begin{tabular}{|c c c|}
${\partial{p_{\rho}}\over{\partial{\rho^\prime}}}$ & ${\partial{p_{\rho}}\over{\partial{z^\prime}}}$ & ${\partial{p_{\rho}}\over{\partial{E}}}$\\
${\partial{p_{z}}\over{\partial{\rho^\prime}}}$ & ${\partial{p_{z}}\over{\partial{z^\prime}}}$ & ${\partial{p_{z}}\over{\partial{E}}}$\\
${\partial{t_{\beta,m}}\over{\partial{\rho^\prime}}}$ & ${\partial{t_{\beta,m}}\over{\partial{z^\prime}}}$ & ${\partial{t_{\beta,m}}\over{\partial{E}}}$.
\end{tabular}

This determinant is solved by using the values of $p_{\rho}, p_{z}$ and $t_{\beta,m}$ from equations (14), (15) and (5), respectively. The explicit expressions are
\begin{equation}
\bigl({\partial{p_{\rho}}\over{\partial{\rho^\prime}}}\bigr)_{{\rho^\prime}\rightarrow{\rho}}={\sqrt{2ME}K_{1}\over{\zeta}}
\end{equation}
where,
\begin{equation}
K_{1}=\bigl[{-T_{1}\over{sin{u}sinh{v}}}+{T_{2}\over{cos{u}cosh{v}}}\bigr]
\end{equation}
\begin{eqnarray}{\nonumber}
T_{1}={\bigl[-Acos{u}sinh{v}+{sin{u}sinh{v}(sin{u}cos{u}+{\sigma_{2}sin{u}\over{cos^3{u}}})\over{A}}-Bsin{u}cosh{v}\bigr]\over{cosh^2{v}-sin^2{u}}}\\
+2sin{u}cos{u}{\bigl[-Asin{u}sinh{v}+Bcos{u}cosh{v}\bigr]\over{(cosh^2{v}-sin^2{u})^2}}
\end{eqnarray}
\nonumber
\begin{eqnarray}{\nonumber}
T_{2}={\bigl[-Asin{u}cosh{v}+{cos{u}cosh{v}(cosh{v}sinh{v}+{\sigma_{2}{cosh{v}}\over{sinh^3{v}}})\over{B}}+Bcos{u}sinh{v}\bigr]\over{cosh^2{v}-sin^2{u}}}\\
-2sinh{v}cosh{v}{\bigl[-Asin{u}sinh{v}+Bcos{u}cosh{v}\bigr]\over{(cosh^2{v}-sin^2{u})^2}}
\end{eqnarray}
\nonumber
\begin{equation}
\big({\partial{p_{\rho}}\over{\partial{z^\prime}}}\big)_{{z^\prime}\rightarrow{z}}={\sqrt{2ME}K_{2}\over{\zeta}}
\end{equation}
where,
\begin{equation}
K_{2}=\bigl[{T_{1}\over{cos{u}cosh{v}}}+{T_{2}\over{sin{u}sinh{v}}}\bigr]
\end{equation}
\begin{equation}
{\partial{p_{\rho}}\over{\partial{E}}}={M\over{\sqrt{2ME}}}K_{3}
\end{equation}
\begin{equation}
K_{3}={\bigl[-Asin{u}sinh{v}+Bcos{u}cosh{v}\bigr]\over{{cosh^2{v}-sin^2{u}}}}
\end{equation}
\begin{equation}
\big({\partial{p_{z}}\over{\partial{\rho^\prime}}}\big)_{{\rho^\prime}\rightarrow{\rho}}={\sqrt{2ME}K_{4}\over{\zeta}}
\end{equation}
where,
\begin{equation}
K_{4}=\bigl[{-T_{3}\over{sin{u}sinh{v}}}+{T_{4}\over{cos{u}cosh{v}}}\bigr]
\end{equation}
\begin{eqnarray}{\nonumber}
T_{3}={\bigl[-Asin{u}cosh{v}-{cos{u}cosh{v}(sin{u}cos{u}+{\sigma_{2}{sin{u}}\over{cos^3{u}}})\over{A}}+Bcos{u}sinh{v}\bigr]\over{cosh^2{v}-sin^2{u}}}\\
+2sin{u}cos{u}{\bigl[Acos{u}cosh{v}+Bsin{u}sinh{v}\bigr]\over{(cosh^2{v}-sin^2{u})^2}}
\end{eqnarray}
\nonumber
\begin{eqnarray}{\nonumber}
T_{4}={\bigl[Acos{u}sinh{v}+{sin{u}sinh{v}(cosh{v}sinh{v}+{\sigma_{2}{cosh{v}}\over{sinh^3{v}}})\over{B}}+Bsin{u}cosh{v}\bigr]\over{cosh^2{v}-sin^2{u}}}\\
-2sinh{v}cosh{v}{\bigl[Acos{u}cosh{v}+Bsin{u}sinh{v}\bigr]\over{(cosh^2{v}-sin^2{u})^2}}
\end{eqnarray}
\nonumber
\begin{equation}
\bigl({\partial{p_{z}}\over{\partial{z^\prime}}}\bigr)_{{z^\prime}\rightarrow{z}}={\sqrt{2ME}K_{5}\over{\zeta}}
\end{equation}
where,
\begin{equation}
K_{5}=\bigl[{T_{3}\over{cos{u}cosh{v}}}+{T_{4}\over{sin{u}sinh{v}}}\bigr]
\end{equation}
\begin{equation}
{\partial{p_{z}}\over{\partial{E}}}={M\over{\sqrt{2ME}}}K_{6}
\end{equation}
\begin{equation}
K_{6}={\bigl[Acos{u}cosh{v}+Bsin{u}sinh{v}\bigr]\over{{cosh^2{v}-sin^2{u}}}}
\end{equation}
\begin{equation}
{\partial{t_{\beta,m}}\over{\partial{\rho^\prime}}}_{{\rho^\prime}\rightarrow{\rho}}={M\over{m\sqrt{2ME}}}K_{7}
\end{equation}
\begin{equation}
K_{7}=\bigl[{-An_{u}\over{sin{u}sinh{v}}}+{Bn_{v}\over{cos{u}cosh{v}}}\bigr]
\end{equation}
\begin{equation}
{\partial{t_{\beta,m}}\over{\partial{z^\prime}}}_{{z^\prime}\rightarrow{z}}={M\over{m\sqrt{2ME}}}K_{8}
\end{equation}
\begin{equation}
K_{8}=\bigl[{An_{u}\over{cos{u}cosh{v}}}+{Bn_{v}\over{sin{u}sinh{v}}}\bigr]
\end{equation}
\begin{equation}
{\partial{t_{\beta,m}}\over{\partial{E}}}=-{M\over{m{2ME}\sqrt{2ME}}}L_{\beta}
\end{equation}
Thus, the above determinant yields
\begin{equation}
J={M^2\over{m\sqrt{2ME}\zeta}}\bigl[-K_{1}K_{5}L_{\beta}-K_{1}K_{8}K_{6}+K_{2}K_{4}L_{\beta}+K_{2}K_{7}K_{6}+K_{4}K_{8}K_{3}-K_{7}K_{5}K_{3}\bigr].
\end{equation}
It may be noted that $J$ diverges for $\zeta$=0.0 or, $\eta$=1. We, therefore, calculate the results for a sphere in the limit $\zeta\rightarrow$0.0 by choosing a deformation $\eta$=1.0001.

\newpage
\begin{table}{Table I: A comparison of eigenvalues, in terms of $kR_{0}$, of Type I, Type II and 3D orbits with that of Quantum mechanical at $\eta=1.3$}
\begin{tabular}{|c|c|c|c|c|c|c|}
N & L & $L_{z}$ & Qmech. & Type I & Type II & 3D\\\hline
1& 0& 0& 3.1927 & 3.45 &- & - \\
1 &1& 0& 4.2817 &- &- & 3.913\\
1& 1& 1& 4.6990 & 4.96&- &- \\
1& 2& 0& 5.072  & - & - & 4.892\\
1& 2& 1 & 5.7211 &- & - & 5.412\\
1& 2& 2 & 6.10   & 6.375& - & -\\
1& 3& 0 & 6.4064 & - & 6.335 &-\\
1&  3&  1& 6.6153 &- &- & 6.352\\
2& 0& 0& 6.8084 & 6.969 &- &-\\
1& 3& 2 &7.0879 & - &- &6.991\\
1& 3& 3& 7.4446 & 7.736 &- & -\\
1& 4& 0& 7.5739 &- & 7.468&-\\
3& 2& 0& 7.6595 &- &- & 7.329\\
1& 4& 1& 7.6926 &- &- & 7.427\\
2 &1& 0& 7 7065 &- &- & 7 737\\
1 &4 &2& 8.0079 &- &- & 8.069\\
2 &1 &1& 8.2915 & 8.575&- &-\\
1& 4 &3 &8.4116 & - & - &8.373\\
1& 5& 0& 8.6986 &- & 8.57&- \\
1&  4&  4&  8.7538 & 9.064&- &- \\
1& 5& 1& 8.7797 & - &- &8.683\\
1 &5 &2& 9.0071 & - &- &9.029\\
2& 2& 1& 9.3079 &- &- & 9.143\\
1 &5 &3& 9.3371 &- &- & 9.358\\
1& 5& 4& 9.7061 & - &- &9.648\\
1 &6& 0& 9.7999 &- & 9.651& -\\
1 &6& 1& 9.8607 & - &- &9.962\\
\end{tabular}
\end{table}
\newpage
\begin{figure}
\caption{Total level density $g(E)$  vs. $kR_{0}$ for Type I orbits at $\eta=1.0001$, and its comparison with quantum results.}
\end{figure}

\begin{figure}
\caption{Total level density $g(E)$  vs. $kR_{0}$ for Type II orbits at $\eta=1.0001$, and its comparison with quantum results.}
\end{figure}
\begin{figure}
\caption{Total level density $g(E)$  vs. $kR_{0}$ for combination of Type I and Type II orbits at $\eta=1.0001$.}
\end{figure}
\begin{figure}
\caption{Total level density $g(E)$ vs. $kR_{0}$ for Type I orbits at $\eta=1.3$, and its comparison with quantum eigenvalues having $L$=$L_z$.}
\end{figure}
\begin{figure}
\caption{Total level density $g(E)$  vs. $kR_{0}$ for Type II orbits at $\eta=1.3$, and its comparison with quantum eigenvalues having $L_z$=0.}
\end{figure}
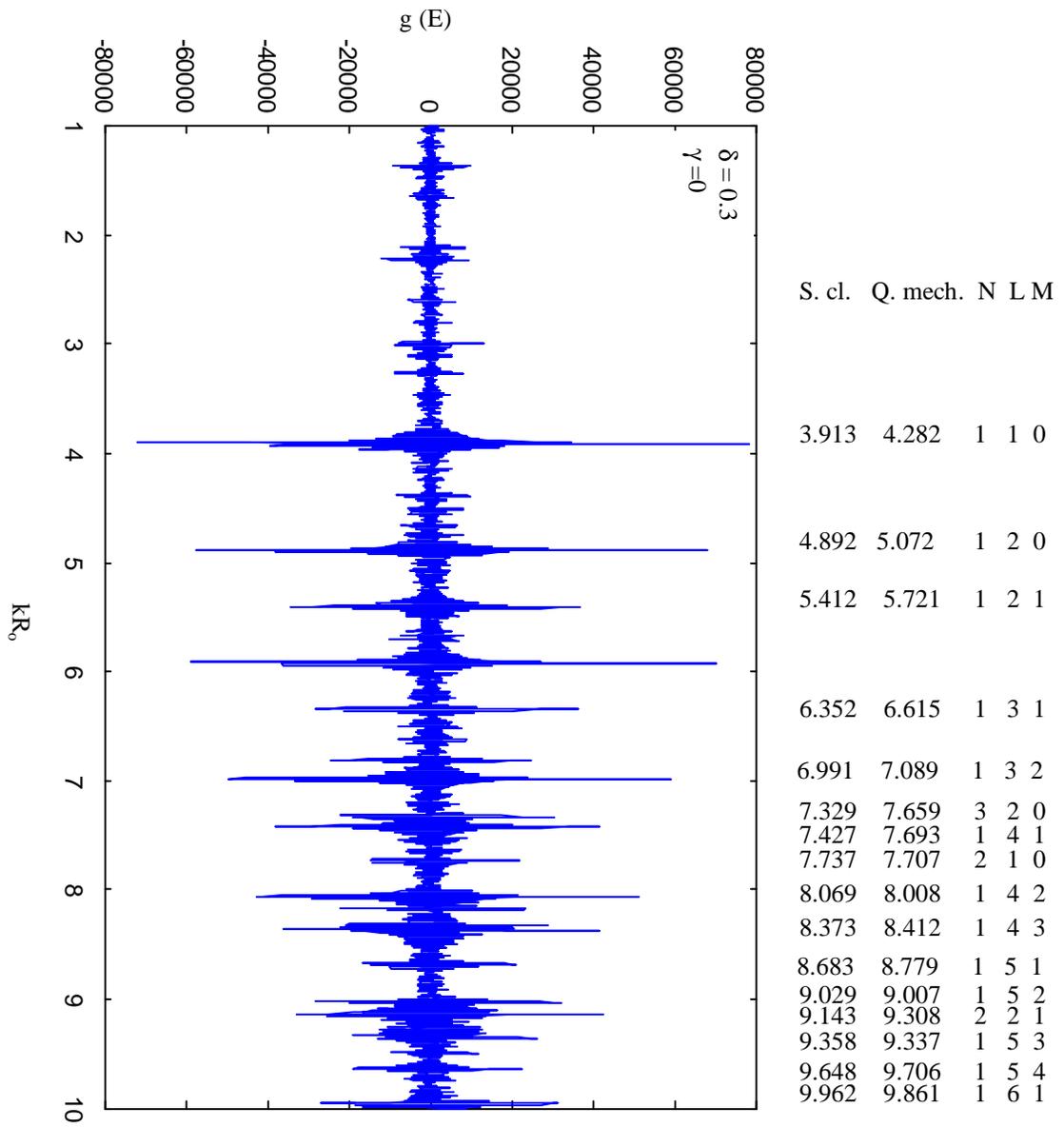
\begin{figure}
\caption{Total level density $g(E)$  vs. $kR_{0}$ for 3D orbits at
$\eta=1.3$, and its comparison with quantum eigenvalues.}
\end{figure}
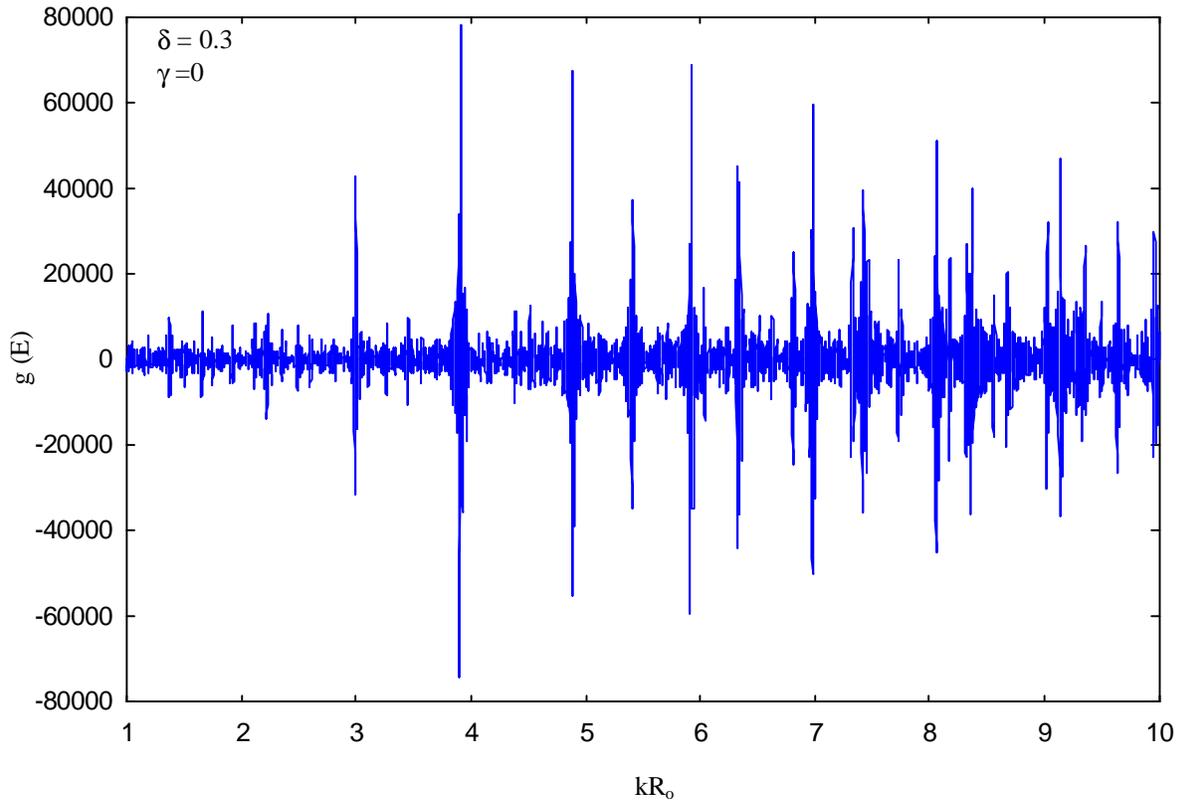
\begin{figure}
\caption{Total level density $g(E)$  vs. $kR_{0}$ for combination
of Type I, Type II and 3D orbits at $\eta=1.3$.}
\end{figure}

\newpage

\end{document}